%% file: imreal-usem.tex
\documentclass{sig-alternate}

\usepackage{amssymb}
\usepackage{amsmath}
\usepackage{graphicx}
\usepackage{url}
\usepackage{subfigure}
\usepackage{booktabs}
\usepackage{multirow}

\title{U-Sem: Semantic Enrichment, User Modeling and Mining of Usage Data on the Social Web} 

\numberofauthors{1} 
\author{
\alignauthor Fabian Abel, Ilknur Celik, Claudia Hauff, Laura Hollink, Geert-Jan Houben\\
       \affaddr{Web Information Systems, TU Delft}\\
       \affaddr{Mekelweg 4, 2628 CD Delft, the Netherlands}\\
       \email{\{f.abel,i.celik,c.hauff,l.hollink,g.j.p.m.houben\}@tudelft.nl}
}

\begin{document}

\maketitle 

\begin{abstract} 
With the growing popularity of Social Web applications, more and more user data is published on the Web everyday. Our research focuses on investigating ways of mining data from such platforms that can be used for modeling users and for semantically augmenting user profiles. This process can enhance adaptation and personalization in various adaptive Web-based systems. In this paper, we present the U-Sem people modeling service, a framework for the semantic enrichment and mining of people's profiles from usage data on the Social Web. We explain the architecture of our people modeling service and describe its application in an adult e-learning context as an example. 
\end{abstract}

\section{Introduction}\label{sec:introduction}
\input{introduction}

\section{U-Sem: People Modeling Service}\label{sec:approach}
\begin{figure}[t!] 
  \centering
  \includegraphics[width=0.25\textwidth]{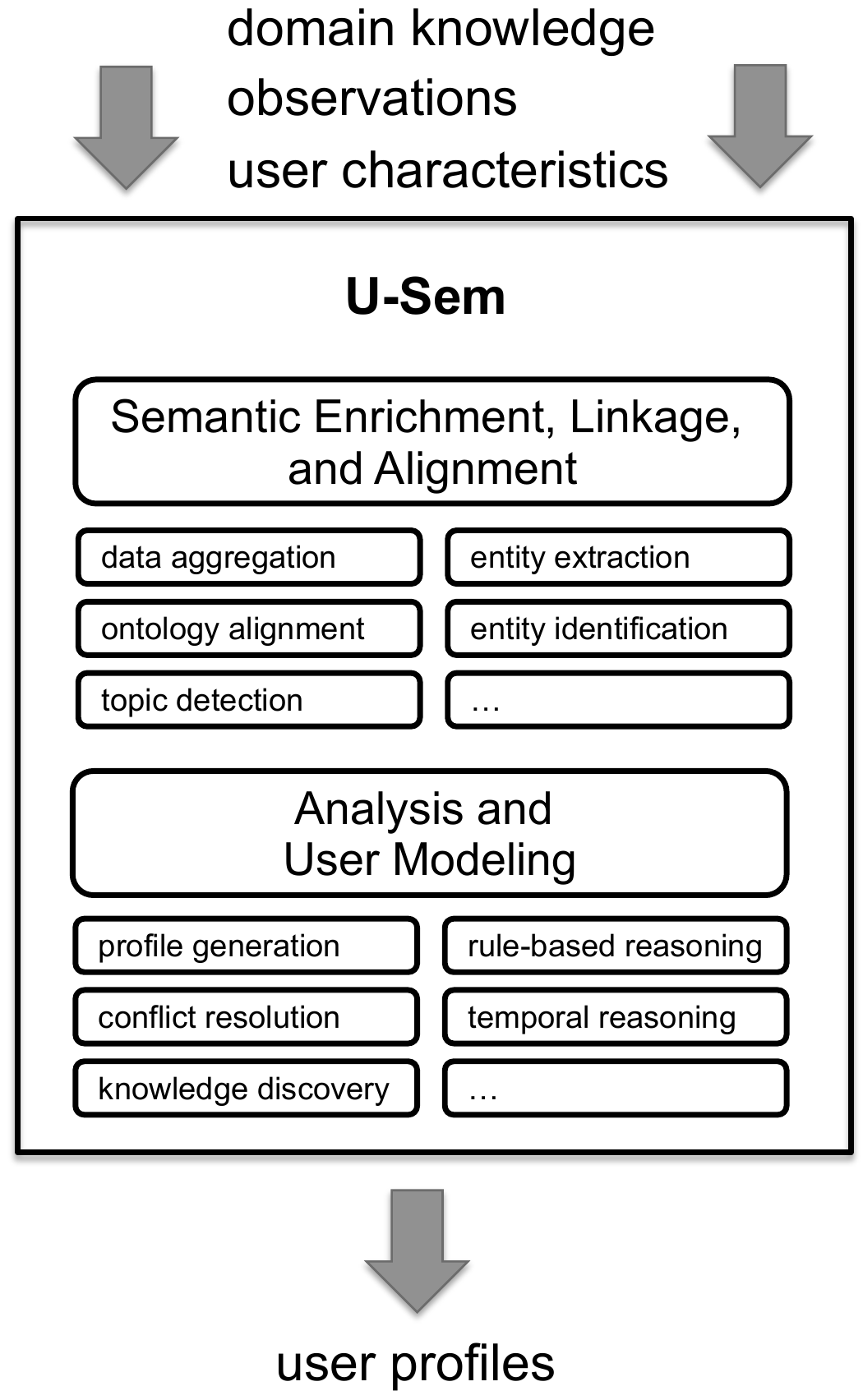}
  \vspace{-0.3cm}
 \caption{Conceptual architecture of the \emph{U-Sem} people modeling service.}
\label{fig:architecture}
\vspace{-0.4cm}
\end{figure}
With the advent of the Semantic Web, generic user modeling formats like GUMO~\cite{HeckmannSB+05UM} or Grapple Core\footnote{http://wis.ewi.tudelft.nl/rdf/grapple-core.owl} and specific profile description vocabularies such as FOAF\footnote{\url{http://xmlns.com/foaf/0.1/}} or SIOC\footnote{\url{http://rdfs.org/sioc/spec/}}, the re-use of usage and user profile data in different application contexts is becoming feasible nowadays. Building on results from previous research on user modeling services~\cite{kobsa2001b,userModeling/personis/2007,gumf/umap/2009}, on mediating user models~\cite{userModelMediation/umuai/BerkovskyKR08} as well as on cross-system user modeling and personalization~\cite{umap2010/profileAggregation,crossSystemPers/UM/Mehta07}, with U-Sem we contribute to a recent line of research which leverages generic user modeling in today's Social Web sphere.

U-Sem is a Semantic Web service for enriching and mining usage and user data.  
Figure~\ref{fig:architecture} shows the architecture of the U-Sem people modeling service.
Given data about people as well as domain knowledge, U-Sem aims to infer and output 
user profiles to support personalization in other applications. We distinguish three types of input data:
\vspace{-0.1cm}
\begin{description}
\item[Observations] By \emph{observation} we refer to usage data or events that give us rather implicit user data such as clicks, tagging activities, bookmarking actions or posts on Twitter. We model observations in RDF, re-using the Grapple Core ontology (namespace abbreviation: \emph{gc}). 
A (\emph{gc:Observation}) has the following properties:
\begin{enumerate}
  \item \emph{gc:user} extends \emph{rdf:subject} and identifies the user who performed an activity that was observed.
  \item \emph{usem:what} extends \emph{gc:predicate}. It refers to the type of activity that was observed (e.g.\ clicked, posted).
  \item \emph{rdf:object} points to the RDF resource that the user interacted with. For example, the resource on which the user clicked or a Twitter message which the user published.
  \item \emph{usem:when} extends \emph{dc:created} from the Dublin Core metadata terms\footnote{\url{http://dublincore.org/documents/dcmi-terms/}} and depicts the time when an observation was done.
  \item Moreover, it is possible to attach further information to an observation such as the URI of the application that did the observation (\emph{gc:creator}) or other provenance information.      
\end{enumerate} 

\item[User characteristics] U-Sem also allows for rather explicit inputs from users about their characteristics. For example, properties like the name, homepage or date of birth can be inputted into U-Sem as well using vocabularies such as FOAF. For selected Social Web services like Facebook or LinkedIn, we also foresee automatic data aggregation modules.
\item[Domain knowledge] In order to infer user profiles that support certain application domains, U-Sem requires domain knowledge. This domain knowledge can be input via plain RDF statements and is to some extent automatically obtained from the Web of Data. 
\end{description}

Given these three types of input data, U-Sem features a variety of plug-ins and components that support
the generation of semantically rich user profiles. These components can be grouped 
into two categories.

\begin{description}
\item[Semantic enrichment, Linkage and Alignment] The semantic enrichment layer provides functionality for aggregating and linking user data from Social Web systems like Facebook or Twitter as well as integration and alignment of RDF data from  Linked Data services such as DBpedia\footnote{\url{http://dbpedia.org/}}. Entity extraction, entity identification and topic detection modules can process text that is referenced from observations, user characteristics or domain knowledge to make semantics more explicit and usable for reasoning modules. For example, given the observation that a user posted a Twitter message, U-Sem aims to link the Twitter message to DBpedia concepts and categorizes the message into broad topics such as sports or politics.

Based on the enriched user and usage data, U-Sem also foresees to enrich the domain knowledge by discovering further knowledge such as additional SKOS relations\footnote{\url{http://www.w3.org/TR/skos-reference/}} that were not yet explicitly specified.
\item[Analysis and User modeling] Given the enriched data, the analysis and user modeling layer allows for generating user profiles that describe interests, knowledge and other characteristics of the users. Therefore, it provides rule-based reasoning as well as data mining functionality and moreover aims to leverage temporal dynamics in the data to, for example, infer current interests of a user or behavioral patterns related to some temporal context. Conflict resolution techniques are required to resolve contradicting statements about users as well as other inconsistencies.    
\end{description}

The user profiles that are delivered by U-Sem describe users' characteristics like names, locations or date of birth 
as well as interests or knowledge regarding certain topics. U-Sem profiles are based on FOAF (namespace: \emph{foaf}) 
and the Weighted Interest Vocabulary\footnote{\url{http://purl.org/ontology/wi/core#}} (namespaces: \emph{wi}, \emph{wo}). Furthermore, we extend these 
vocabularies to express users' knowledge or opinion regarding specific concepts (namespace: \emph{usem}). For example, 
the following RDF/Turtle\footnote{\url{http://www.w3.org/TeamSubmission/turtle/}} representation specifies to what degree a user (\emph{http://bob.my\-openid.com}) has knowledge 
about a certain topic (\emph{dbpedia: Psychology}). The meaning of the weight (\emph{wo:weight\_value}) is
defined via a scale (\emph{ex:AScale}) that is referenced from the \emph{usem:WeightedKnowledge} instance.

\vspace{-0.2cm}
\begin{scriptsize}
\begin{verbatim}
<http://bob.myopenid.com>
   a foaf:Person ;
   foaf:name "Bob";
   usem:knowledge [
      a usem:WeightedKnowledge ;
      wi:topic dbpedia:Psychology ;
      wo:weight [
         a wo:Weight ;
         wo:weight_value 10.0 ;
         wo:scale ex:AScale
         ]  .
\end{verbatim}
\end{scriptsize}
\vspace{-0.2cm}

From a technical point of view, U-Sem accepts data in RDF format and allows for querying by means of SPARQL\footnote{\url{http://www.w3.org/TR/rdf-sparql-query/}}. To facilitate the retrieval of certain types of profiles, we plan to complement SPARQL with further operators. For example,
as U-Sem aggregates user data from different Social Web services, there is a demand for operators that allow to specify how user data should be combined to prioritize certain sources and resolve possible conflicts.



\section{Application Scenario}\label{sec:application-scenario}
Given the architecture described in the previous section, U-Sem coordinates input from various sources to enrich user profiles as well as domain knowledge for enhancing personalization. In this section, we present an example scenario within the context of ImREAL, where the U-Sem framework will be used with a training simulator in the medical domain. The flow of data will be as follows: this simulator gathers a set of basic user characteristics and click data that are fed to the U-Sem framework together with domain information; U-Sem performs semantic enrichment and user modeling via a number of modules in order to provide enriched user profiles and domain knowledge, which are fed back to the simulator for improved adaptation.  

\subsection{Semantic Enrichment}\label{sec:application-scenario-enrichment}
U-Sem adds context and semantics to the observations (e.g.\ click data) made in an application, in order to infer user knowledge levels or interests for a given topic. In the context of ImREAL U-Sem may, for example, retrieve
the following click observation. 

\vspace{-0.2cm}
\begin{scriptsize}
\begin{verbatim}
<http://imreal-project.eu/observation/1>
  gc:user <http://bob.myopenid.com>;
  gc:predicate imreal:accessed;
  gc:object <http://imreal-project.eu/resource/1234567>
  gc:created "2011-02-15 20:10:30" .
\end{verbatim}
\end{scriptsize}
\vspace{-0.2cm}

Given this observation, U-Sem extracts and identifies concepts that are mentioned in or related to the resource (\emph{http:/ /imreal-project.eu/resource/1234567}) the user accessed. 
The U-Sem topic detection module allows for automatic classification of resources.
The concepts and topics extracted from click data are also linked to related Linked Data vocabularies, such as MeSH\footnote{\url{http://www.ncbi.nlm.nih.gov/mesh}} and DBpedia. Such linkage makes the semantic meaning of observations explicit and aligns the knowledge embodied in observations that are possibly done in different applications. For example, the above resource may be linked to \emph{dbpedia:Psychology} and may allow the analysis and user modeling layer to infer Bob's knowledge level regarding this concept (see the example user profile listed in Section~\ref{sec:approach}).

The U-Sem data aggregation modules (see Figure~\ref{fig:architecture}) allow for automatic integration of user characteristics and user activities from selected Social Web applications. For instance, U-Sem exploits social networking services such as LinkedIn or Facebook to gather user characteristics, social bookmarking services such as 
CiteULike
to monitor documents that people are interested in and blogging services such as Twitter to complement information about users with their opinions and thoughts on certain topics. For example, U-Sem may retrieve the following three observations from LinkedIn, CiteULike and Twitter respectively: 

\vspace{-0.2cm}
\begin{scriptsize}
\begin{verbatim}
<http://linkedin.com/bob>
   a foaf:Person ;
   foaf:name "Bob";
   foaf:workplaceHomepage <http://mh-hannover.de/> .

<http://imreal-project.eu/observation/3>
  gc:user <http://citeulike.org/bob>;
  gc:predicate imreal:bookmarked;
  gc:object <http://www.citeulike.org/article/67893712>
  gc:created "2011-02-15 22:05:00" .

<http://imreal-project.eu/observation/4>
  gc:user <http://twitter.com/bob>;
  gc:predicate imreal:posted;
  gc:object <http://imreal-project.eu/resource/twitter/1234567>
  gc:created "2011-02-15 22:00:00" .
  
<http://imreal-project.eu/resource/twitter/1234567>
  tw:id 1234567;
  tw:username <http://twitter.com/bob>
  tw:content "Interesting article on Chopin's #hallucinations: 
              http://bit.ly/y4Gfs5";
  tw:creationTime "2011-02-15 21:45:00" .
\end{verbatim}
\end{scriptsize}
\vspace{-0.2cm}

Information about the users' current and previous jobs, interests or knowledge in certain topics, and professional groups, can be extracted from their public LinkedIn profiles. Extracting a user's professional web pages is shown above as an example. Such observations are then used to supplement the given basic user characteristics. 

Similarly, users' professional interests and knowledge can be extracted from services like CiteULike where users tag articles as shown in the example above. The articles (\emph{http://ww w.citeulike.org/article/67893712}) are analyzed by the topic detection module of U-Sem to extract the concepts covered, such as ``diagnosis", ``symptoms", ``psychosis" and ``mania", which are then linked to rich vocabularies such as DBpedia and MeSH for enrichment. We will also investigate if and how we can infer users' knowledge of related concepts by traversing the nodes of the vocabulary or by using a semantic similarity measure such as~\cite{Anand:2007:GSE:1278366.1278371}.

Additionally, U-Sem employs modules to monitor user activities in popular real-time microblogging sites like Twitter as given in the example observation above. Thus, when the user posts a tweet 
(\emph{http://imreal-project.eu/resource/twitter /1234567}), 
the entity identification module identifies and extracts entities both from the content of the tweet (\emph{``Interes-ting article on Chopin's \#hallucinations: {http://bit.ly/y4Gfs 5}"}) and from the destination of the URL referenced in the content (\emph{http://bit.ly/y4Gfs5}). These are then linked to the corresponding concepts in, for example, DBpedia, and appended to the set of user characteristics as weighted interest of the user in the given concepts. Other user actions such as bookmarking in Delicious or tagging in Flickr are treated as observations in a similar manner where extracted entities are assigned weighted knowledge and interest levels, and then attached to the set of user characteristics.

\subsection{User Modeling}\label{sec:application-scenario-um}
U-Sem utilizes a variety of modules to reason, infer, generate and aggregate user profiles from the available enriched usage and user data. For instance, given the Twitter observation before, the reasoning module considers all entities extracted both from the text of the Twitter message and the content of the URL mentioned in the tweet. The module deduces that the user is interested in ``Frederic Chopin" and reads about ``Hallucinations" and ``Epilepsy". Based on the fact that both concepts, ``Hallucinations" and ``Epilepsy", are in the ``diseases" category in the DBpedia hierarchy, this module can reason that this user has knowledge about diseases and can adjust the knowledge level in these concepts accordingly. 

Since U-Sem gathers data from different sources, a dedicated component is required to oversee data aggregation and possible conflict resolutions. When similar user characteristics are extracted from different sources with conflicting values, this module decides on how to determine the final value depending on the reliability of the sources, by checking the internally assigned trust values of the sources, and the timestamps of the observations. For instance, during the data aggregation phase, U-Sem receives two observations for a user: one click data observation which mentions the user answered a question related to ``diagnosing psychotic symptoms" incorrectly, and another one from CiteULike two hours later which states that the user tagged an article on ``The art of distinguishing anxiety and psychotic symptoms in diagnosing Mania'', where the user identification module maps (\emph{http://citeulike.org/\-bob}) and (\emph{http://bob.my\-openid.com}) to the same user, \emph{Bob}. Although the usage data suggests that the user's knowledge levels in ``diagnosis", ``symptoms", ``psychotic" and ``mania" are low, the user's CiteULike activity shows that the user recently read up on these concepts and therefore the knowledge levels in these concepts should be increased.

The final step in this application scenario would be to feed back to the simulator application all the above mentioned extracted and inferred user characteristics and interest/knowledge levels, represented as a (extended) FOAF profile. For example, \emph{Bob's} user profile given at the end of Section~\ref{sec:approach} would be extended to include also his  \emph{usem:Weighted\-Knowledge} of \emph{dbpedia:Hallucination}, \emph{dbpedia:Epilepsy} and \emph{dbpedia:Dis\-eases} in general, as well as user characteristics such as his work home page extracted from LinkedIn.

\vspace{-0.15cm}
\section{Outlook}\label{sec:conclusion}
\input{conclusion}

\vspace{0.30cm}
\textbf{Acknowledgements.} This work is partially sponsored by the EU FP7 project ImREAL (http://imreal-project.eu).
\vspace{-0.45cm}

\bibliographystyle{abbrv}


\end{document}

%% file: introduction.tex

Social Web stands for a new culture of participation on the Web. In the last decade, people got more and more involved into contributing and shaping content on the Web. People share their thoughts on microblogging systems like Twitter, profile themselves on websites like LinkedIn, publish their bookmarks on Delicious or use services like CiteULike to organize scientific publications they are interested in\footnote{\url{http://{twitter,linkedin,delicious,citeulike}.com}}. Hence, they leave a variety of data traces on the Web~\cite{umap2010/profileAggregation}. Exploiting these traces promises to be very beneficial for various applications that aim for adaptation and personalization~\cite{adaptiveWeb/2007}.  

In this paper, we present ongoing work on the U-Sem people modeling service, a framework for the semantic enrichment and mining of people's profiles from usage data on the Social Web. We explain the architecture of our people modeling service and give examples of how U-Sem will be applied to one particular domain, namely e-learning, in the context of the European project ImREAL\footnote{\url{http://www.imreal-project.eu/}}. Platforms that facilitate e-learning have become increasingly prevalent in recent years. Web-based e-learning systems are particularly attractive, as they allow learning at any time, at any place and at any pace. One aspect in e-learning that is not sufficiently addressed yet is the ability of systems to adapt to each individual learner~\cite{userModels/adaptiveHypermedia/Brusilovsky/2007,adaptiveHypermedia/frameworks/debra/1999}. Learners who already know a part of the curriculum can easily become bored if the system does not take their knowledge level into account and adapts the learning units appropriately. On the other end of the spectrum, a system should be able to recognize if a learner is overwhelmed with the curriculum and adapt the learning units accordingly, for example, by offering additional introductory material. E-learning environments, while also being increasingly introduced at schools and universities to aid classroom learning, are especially well suited for adults who often have very limited amount of time. In adult learning systems that aim to teach skills relevant to the work place, the learners' individual professional tasks should also be taken into account, such that the system can place particular emphasis on aspects that are relevant to their work. The overall goal of ImREAL is to improve virtual training of adults by aligning the learning experience in the virtual environment with real-world context and real-world job-related experiences.

Adaptation and personalization of an e-learning system requires user profiling, that is, the collection of data and information about a user~\cite{userModels/adaptiveHypermedia/Brusilovsky/2007}. This process can either be performed explicitly by asking the users a series of questions about their knowledge levels, skills, etc. or implicitly by deriving a user profile from already existing data. In the ImREAL project, we focus on the latter case, specifically, we investigate how to derive user profiles for adult e-learning from the Social Web. The motivation for this work stems from the fact that Social Web services such as Twitter, Delicious or CiteULike are becoming increasingly popular. It is not unlikely that a user of an e-learning system is using one or even several of these Social Web services. Moreover, since this type of data is publicly available on the Web, there will be few privacy concerns for the users. Consequently, we only require the learner's username in one or more Social Web services as input to the user profiling module; only publicly accessible data will be utilized to build the profile.

This general idea leads to several research questions that we would like to answer with U-Sem, namely:
\begin{itemize}
\item How can user profiles, in particular profiles describing interests and knowledge levels of users, be built from Social Web data? 
\item Which Social Web services are most suitable to derive user profiles for various application domains such as e-learning or recommender systems?
\item How can data of several Social Web services be aggregated to generate better user profiles?
\end{itemize}

The focus of our work is on the question of how to extract user profiles from Social Web data. The problem here is the unstructured nature of the Social Web data as well as the fact that users do not only tweet or bookmark Web pages related to their work and professional knowledge, but that a large amount of data is often about their social life, their interests or focused on current news stories. While we focus on the e-learning domain as part of the ImREAL project, we note that U-Sem has a much wider applicability and is not restricted to this particular domain.


%% file: conclusion.tex
In this paper, we have introduced U-Sem, a generic framework for enriching, modeling and mining profile and usage data available on the Web. U-Sem supports Web-based systems that require user profile information and aim for personalization. In particular, we have shown examples of how we envision it to function in the context of e-learning. 
Currently, we investigate the extraction of user profiles from Twitter that model the users' interests and evaluate them in the context of news recommender systems. This line of work has already resulted in the development of the entity extraction, the entity identification and the topic detection modules. We have also started to conduct first experiments in the extraction of user profiles that model the users' knowledge levels. Here, we rely on social bookmarking services that are designed for scholarly works, specifically CiteULike and Bibsonomy.
The next steps of our work will focus on the exploitation of observations from applications (click data, potentially eye tracking) for the user model as well as on the question of how to semantically enrich these observations depending on the demanded types of profiles.